\pdfoutput=1
\documentclass[%
 reprint,
 amsmath,amssymb,
 aps,
]{revtex4-1}

\usepackage{graphicx}
\usepackage{dcolumn}
\usepackage{bm}
\usepackage{color}

\begin{document}

\preprint{APS/123-QED}

\title{The $^{23}$Na(\boldsymbol{$\alpha$},p)$^{26}$Mg reaction rate at astrophysically relevant energies}

\author{A.M. Howard}
 \email{alan.howard@phys.au.dk}
\author{M. Munch}
\author{H.O.U. Fynbo}
\author{O.S. Kirsebom}
\author{K.L. Laursen}
\affiliation{
	Department of Physics and Astronomy, Aarhus University, 8000 Aarhus C, Denmark\\
}

\author{{C.Aa.~Diget}}
\author{N.J.~Hubbard}
\affiliation{
	Department of Physics, University of York, York Y010 5DD, United Kingdom\\
}

\begin{abstract}
The production of $^{26}$Al in massive stars is sensitive to the $^{23}$Na($\alpha$,p)$^{26}$Mg cross section. Recent experimental data suggest the currently recommended cross sections are underestimated by a factor of $\sim$40. We present here differential cross sections for the $^{23}$Na($\alpha$,p)$^{26}$Mg reaction measured in the energy range E$_{\text{c.m.}}=1.7-2.5$~MeV. Concurrent measurements of Rutherford scattering provide absolute normalisations which are independent of variations in target properties. Angular distributions were measured for both p$_{0}$ and p$_{1}$ permitting the determination of total cross sections. The results show no significant deviation from the statistical model calculations upon which the recommended rates are based. We therefore retain the previous recommendation without the increase in cross section and resulting stellar reaction rates of a factor of 40, impacting on the $^{26}$Al yield from massive stars by more than a factor of three.
\end{abstract}

\pacs{25.55.-e}
\maketitle

The observation of $^{26}$Al in the galactic medium, through $\gamma$-ray emission from its daughter nucleus $^{26}$Mg, provided direct evidence for ongoing nucleosynthesis in the galaxy \cite{1982ApJ...262..742M}. While the origins of $^{26}$Al remain the subject of discussion, the C/Ne convective shell within massive stars is a candidate site \cite{1978ApJ...224L.139A}. A sensitivity study of the reactions influencing $^{26}$Al production in massive stars has indicated a significant dependence on the $^{23}$Na($\alpha$,$p$)$^{26}$Mg reaction rate, which acts as a proton source for the $^{25}$Mg(p,$\gamma$)$^{26}$Al reaction \cite{0067-0049-193-1-16}. Specifically, it was found that an increase in the $^{23}$Na($\alpha$,$p$)$^{26}$Mg rate by a factor of 10 would lead to an increase in $^{26}$Al production by a factor of 3.

The $^{23}$Na($\alpha$,$p$)$^{26}$Mg rate adopted in Ref.~\cite{0067-0049-193-1-16} is obtained from statistical model calculations. While earlier experimental data do exist \cite{Kuperus19642253, PhysRevC.9.996}, these were excluded due to a lack of understanding of the target properties during the intense beam bombardment. As a consequence there are significant uncertainties in the experimentally determined resonance strengths.

A recent direct measurement of the reaction cross section in inverse kinematics was made to resolve these experimental uncertainties \cite{PhysRevLett.112.152701}. A $^{23}$Na beam was incident on a gas cell containing $^4$He and outgoing protons corresponding to the ground and first-excited states in $^{26}$Mg detected. The cross sections measured in the region E$_{\text{c.m.}}$~=~$1.7$--$2.5$~MeV were $\sim$40 times greater than statistical model calculations. Such an increase is significantly larger than that required to alter the production of $^{26}$Al by a factor of 3.

A similar, although less dramatic, disagreement with statistical model calculations has been reported for the $^{33}$S($\alpha$,$p$)$^{36}$Cl reaction \cite{PhysRevC.88.065802}. It is noted in Ref.~\cite{PhysRevC.89.058801} that the measured cross sections significantly exceed the expected single particle strength and that, in light of the $^{23}$Na($\alpha$,$p$)$^{26}$Mg results also, there is an urgent need for additional ($\alpha$,$p$) data in the 20~$\leq$~A~$\leq$~50 region.

In this letter we report on a new measurement of the $^{23}$Na($\alpha$,$p$) cross section in forward kinematics covering the energy range E$_{\text{c.m.}}$~=~$1.7$--$2.5$~MeV. Our methodology exploits the simultaneous detection of Rutherford scattered $\alpha$ particles to remove dependencies on properties of the target, such as thickness and stoichiometry, which has impacted previous measurements. Discussions of this methodology may be found in, for example, Refs.~\cite{PhysRevC.65.064609, PhysRevC.77.035801}.

\begin{figure}[h]
\includegraphics[width=75mm]{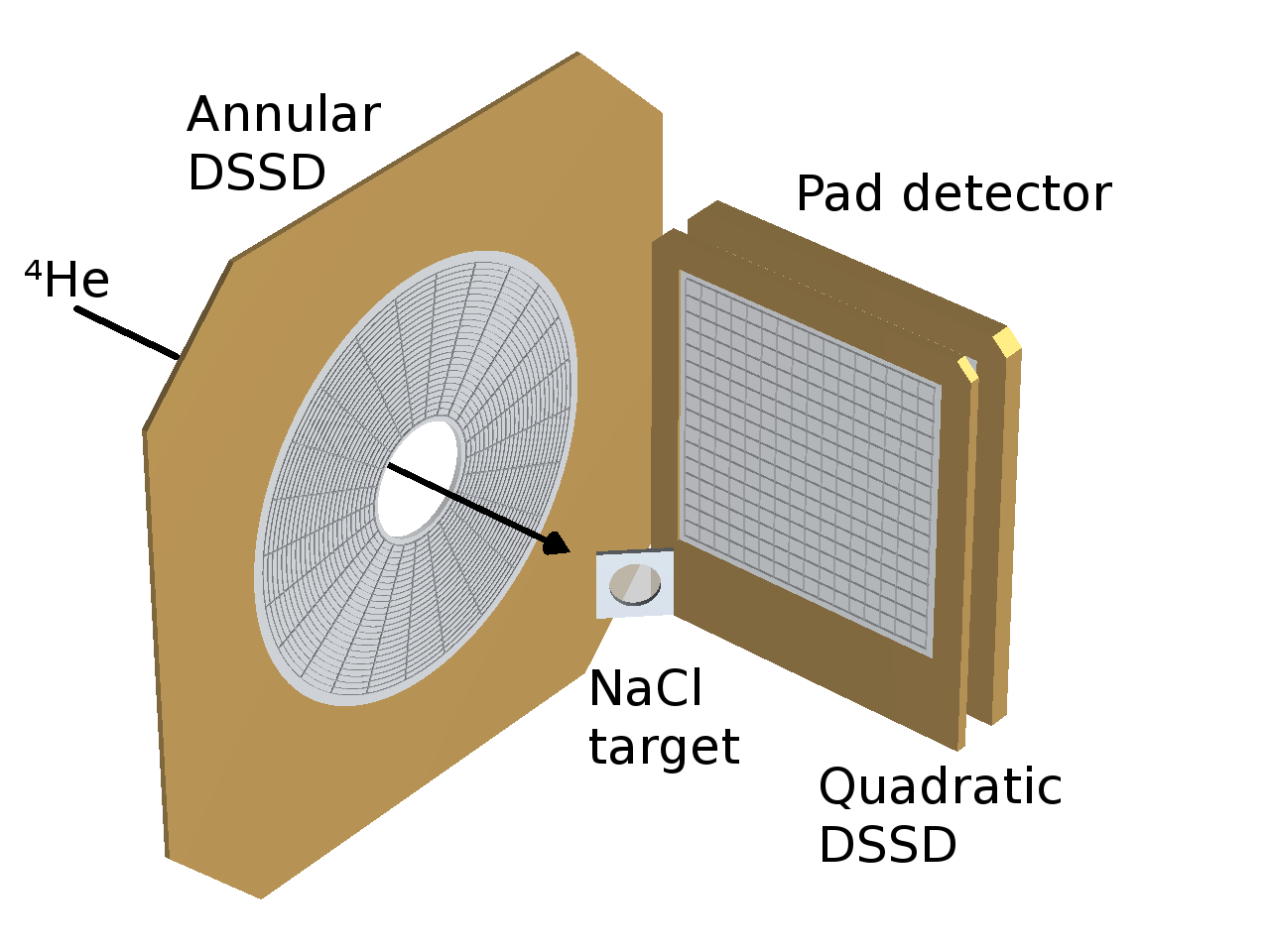}
\caption{\label{fig:schematic} A schematic of the experimental setup within the scattering chamber. The incoming $^4$He beam is indicated by the arrow. The NaCl target was orientated at 45$^{\circ}$ relative to the beam axis. Two double-sided silicon strip detectors were used to detect outgoing protons and $\alpha$ particles, see the text for details. For clarity both front- and back-side segmentation of the detectors is shown.}
\end{figure}

Measurements were made at the Aarhus University 5-MV Van de Graaff accelerator. A schematic of the experimental setup is shown in Fig. \ref{fig:schematic}. A $^4$He beam with laboratory energy between 1.99 and 2.94~MeV was used to bombard a carbon-backed NaCl target. The beam was stopped 70-cm downstream of the target position in a suppressed Faraday cup connected to a current integrator. Typical beam currents were in the range 200--500~ppA.

Two double-sided silicon strip detectors (DSSDs) were mounted in the scattering chamber to provide energy and angle information for outgoing charged particles. A 322-$\mu$m annular DSSD was mounted upstream of the target, covering laboratory angles between 140$^{\circ}$ and 163$^{\circ}$ and a 40-$\mu$m-thick, quadratic DSSD provided coverage at laboratory angles between 60$^{\circ}$ and 120$^{\circ}$. The annular detector was mounted with the junction side, which has a 4$\mu$m dead layer, facing the target. In this orientation the dead layer acts as a degrader foil, increasing the energy separation between backscattered $\alpha$ particles and protons populating $^{26}$Mg. Protons populating the ground and first excited states in $^{26}$Mg were sufficiently energetic to punch through the quadratic DSSD and were stopped, and unambiguously identified, in a 1500-$\mu$m silicon pad detector.

The target was prepared at Aarhus University by evaporating NaCl onto a 10~$\mu$g/cm$^{2}$ carbon foil. The beam energy-loss in the target was calibrated using alpha particles backscattered from the carbon backing into the annular DSSD. As the target is rotated through 180$^{\circ}$ the energies are shifted due to losses within the NaCl layer (see, for example, Ref.~\cite{Chiari2001309} for details of this technique). A thickness of 65~keV at a beam energy of 3~MeV was determined. It should be noted that the target was tilted at 45$^{\circ}$ to the beam axis during all other measurements giving an effective thickness of between 92 and 115~keV for the range of beam energies used.

\begin{figure}[h]
\includegraphics[width=80mm]{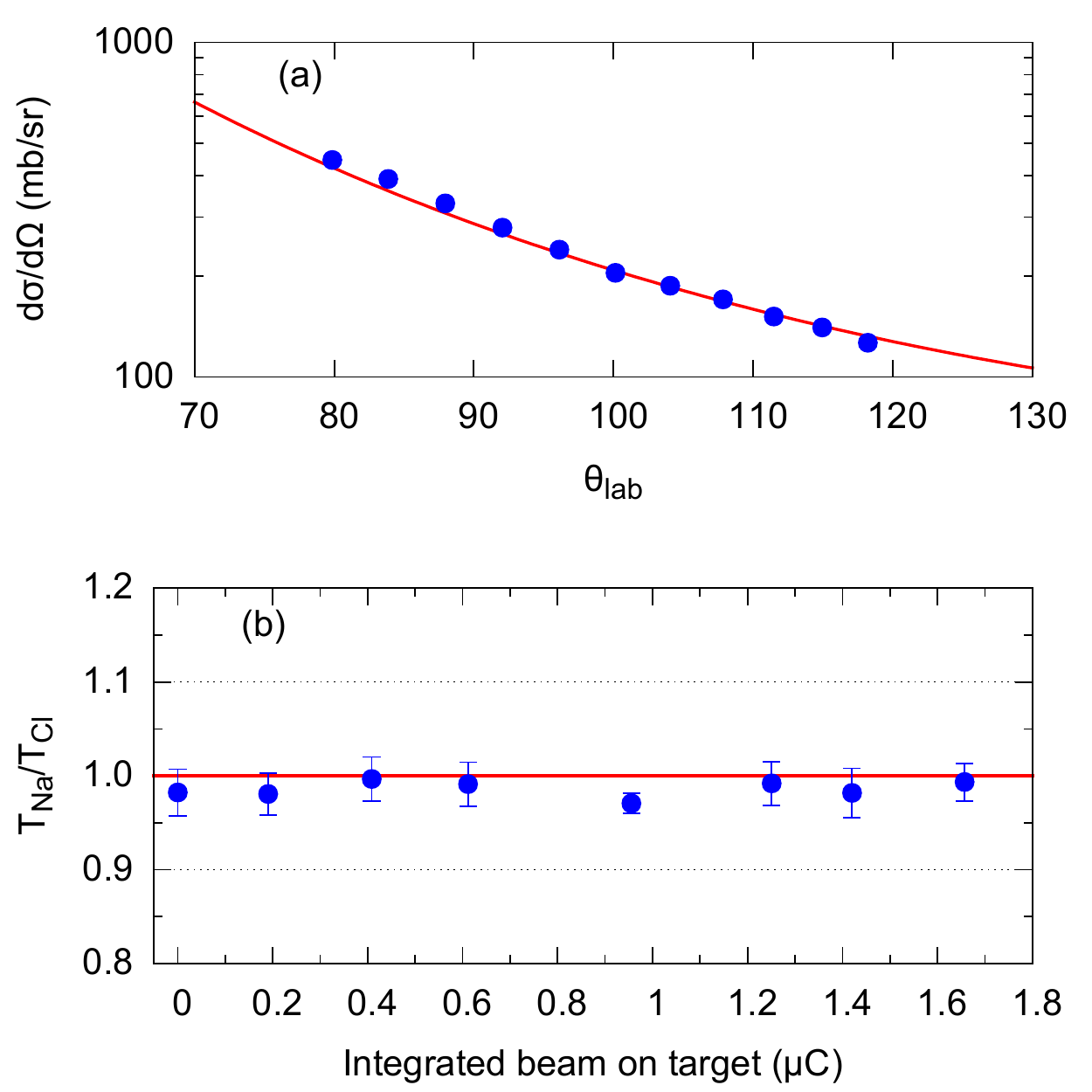}
\caption{\label{fig:target} (a) The measured angular distribution of elastically scattered alpha particles from $^{23}$Na at a beam energy of 2.94~MeV. The solid line is the distribution expected for pure Rutherford scattering. (b) The stoichiometric ratio of Na and Cl in the target foil, as determined by Rutherford scattering, as a function of the integrated beam impinging on the foil.}
\end{figure}

During the experiment elastically scattered alpha particles were continuously measured in the quadratic DSSD. For pure Rutherford scattering the elastic yield is a product of the target thickness and incident beam current. This removes any uncertainties due to changes in the target thickness or stoichiometry, in addition to uncertainties in the integration of beam charge. The $\alpha$ scattering data presented in Ref.~\cite{Cheng1991749} demonstrate that elastic scattering from Na is well described by the Rutherford formula for beam energies up to 3~MeV, which covers the entire range of measurements here. This is supported by a measurement of the angular distribution for elastically scattered $\alpha$ particles from Na measured with our setup, shown in Fig.~\ref{fig:target}, which shows excellent agreement with Rutherford scattering.

Rutherford scattering data were also collected for Cl throughout the experiment. A comparison between the relative amounts of Na and Cl in the target shows no deviation from a ratio of 1:1 to within 10\% for the duration of the experiment, see Fig. \ref{fig:target}. Repeat measurements of the target thickness were also consistent, indicating no significant changes in the target properties during the experiment. This is not surprising given the relatively low beam currents employed, three orders of magnitude lower than those used in Ref.~\cite{PhysRevC.9.996} where significant target degradation was observed.

\begin{figure}[h]
\includegraphics[width=80mm]{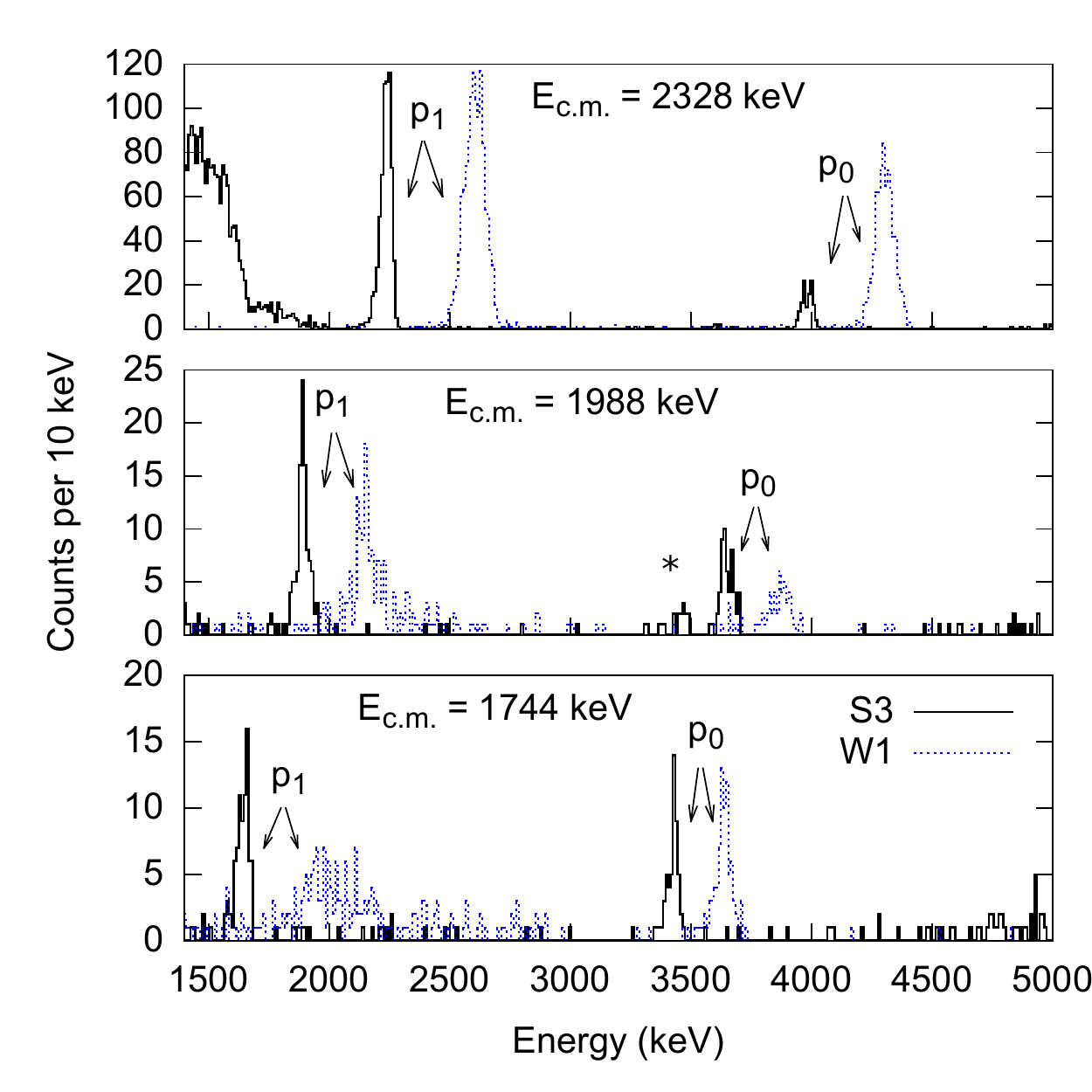}
\caption{\label{fig:spectra} Representative energy spectra from the annular DSSD (S3) and the quadratic DSSD plus pad detector telescope (W1). In the latter case a coincidence between the two detectors is required to remove the background due to $\alpha$ particles stopping in the quadratic DSSD. The effective centre of mass energy in each case is given in the plot, see the text for details. The small peak at $\boldsymbol{^{\ast}}$ has an energy consistent with the $p_0$ transition in $^{19}$F($\alpha$,$p$)$^{22}$Ne, and may therefore be indicative of a thin layer of $^{19}$F on the target surface, see the text for further discussion.}
\end{figure}

Energy spectra for the annular DSSD and quadratic DSSD plus pad detector telescope are shown in Fig.~\ref{fig:spectra}. Proton yields were extracted for both $p_0$ and $p_1$ transitions across the full energy range covered. Differential cross sections were obtained using the normalisation provided by Rutherford scattering of $\alpha$ particles into the quadratic DSSD.

\begin{figure}[h]
\includegraphics[width=80mm]{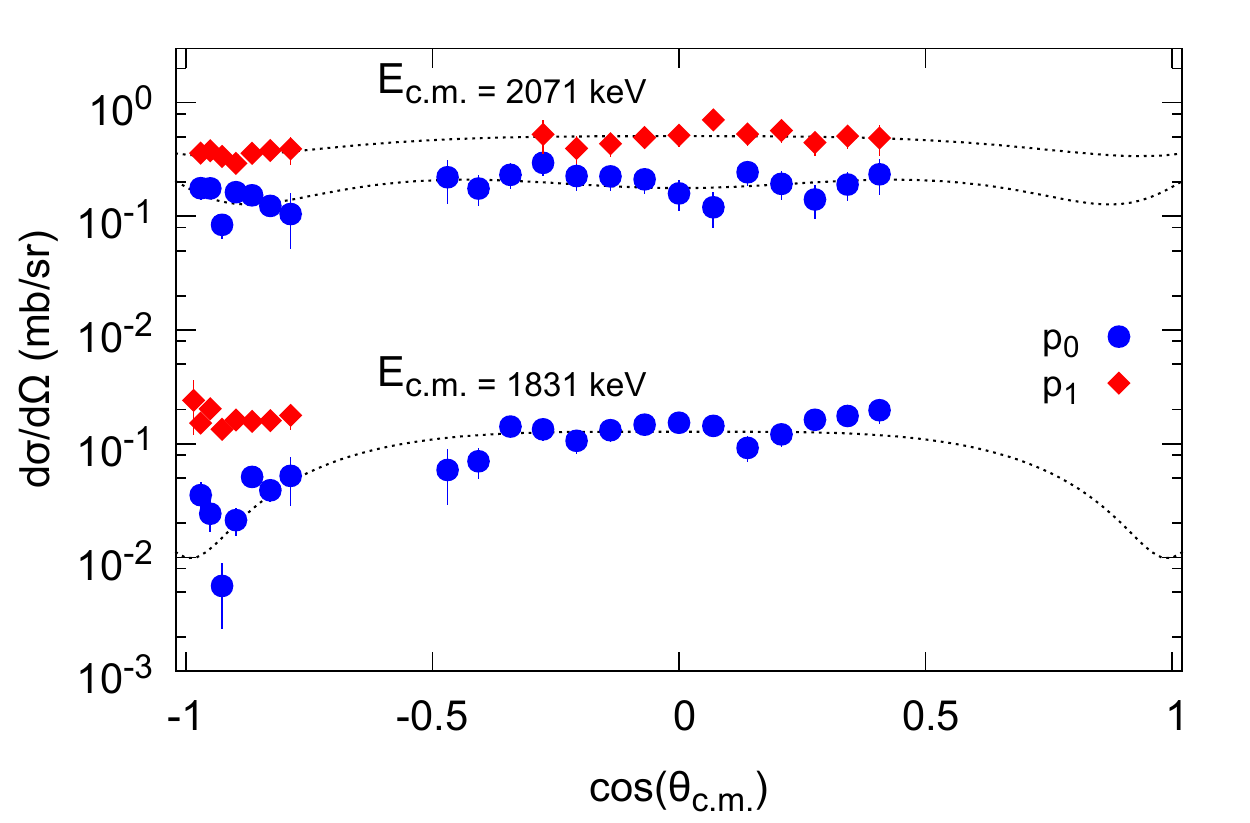}
\caption{\label{fig:distribution} Proton angular distributions from the $^{23}$Na($\alpha$,$p$)$^{26}$Mg reaction. The energies given are the effective centre of mass energies, corrected for the target thickness. For the lowest energy $p_1$ data shown reliable differential cross sections could only be obtained in the annular DSSD detector, located at backward angles in the laboratory frame. The dotted lines show fits of Legendre polynomials to the data.}
\end{figure}

Examples of measured angular distributions are shown in Fig.~\ref{fig:distribution}. To permit total cross sections to be determined measured differential cross sections were fitted using a sum of even-termed Legendre polynomials. This assumes a distribution symmetric around $\theta_{\text{c.m.}}=90^{\circ}$, which is expected when the cross section is dominated by compound nucleus formation. In the measurements reported in Ref.~\cite{Kuperus19642253} only a single, relatively minor resonance was found to exhibit forward-backward asymmetry in the energy region covered here. Nonetheless, a conservative 20\% uncertainty on the total cross section is assumed.

At beam energies below $\sim$2.2~MeV, some fraction of $p_1$ protons reach the pad detector with insufficient energy to be registered. The result is a decrease in the detection efficiency which is not easily quantified. For the two data sets collected below this energy the $p_1$ data collected using the pad detector are therefore not used for the fitting of angular distributions. Instead only data from the annular DSSD, which suffers no decrease in detection efficiency, is used and an isotropic angular distribution assumed. Applying the same procedure to the $p_1$ data sets at higher energy results in a decrease in total cross section of between 10\% and 30\%. A 30\% uncertainty is assumed for the two lowest $p_1$ data points to reflect this.

In two of the eight measurements a weak peak was observed $\sim 200$~keV in energy below $p_0$ (see the middle panel of Fig.~\ref{fig:spectra}). This peak may be indicative of a thin layer of fluorine on the target surface since it lies at the approximate energy expected for $^{19}$F($\alpha$,$p_0$)$^{22}$Ne. Under these circumstances there may be a contribution from $^{19}$F($\alpha$,$p_1$)$^{22}$Ne also, which would not be resolved from $^{23}$Na($\alpha$,$p_1$)$^{26}$Mg. The only possible effect of this could be to increase the observed cross section. Based upon the data presented in Ref. \cite{Kuperus19651603} it is estimated that this contribution should always be below 10\% of the total peak yield. Out of caution an additional 10\% uncertainty is therefore assumed on the lower bound of all $p_1$ cross sections.

Protons populating higher lying states in $^{26}$Mg were not observed in this work due to the background from scattered beam. The contribution from these states to the total reaction cross section at the energies measured in this work is expected to be minor due to the reduced penetrabilities. In Ref.~\cite{PhysRevLett.112.152701} a Hauser-Feshbach calculation is reported which indicates negligible contribution from $p_2$ within the Gamow window (E$_{\text{c.m.}}\simeq1.2$--$2.2$~MeV).

To account for energy losses within the NaCl layer of the target, the measured cross sections are associated with an effective beam energy. This is calculated using an energy dependence for the cross section as given by the statistical model code Non Smoker \cite{RAUSCHER200147}. The resulting effective energies are within 15~keV of the beam energy at the target mid-point for all measurements.

\begin{figure}[h]
\includegraphics[width=80mm]{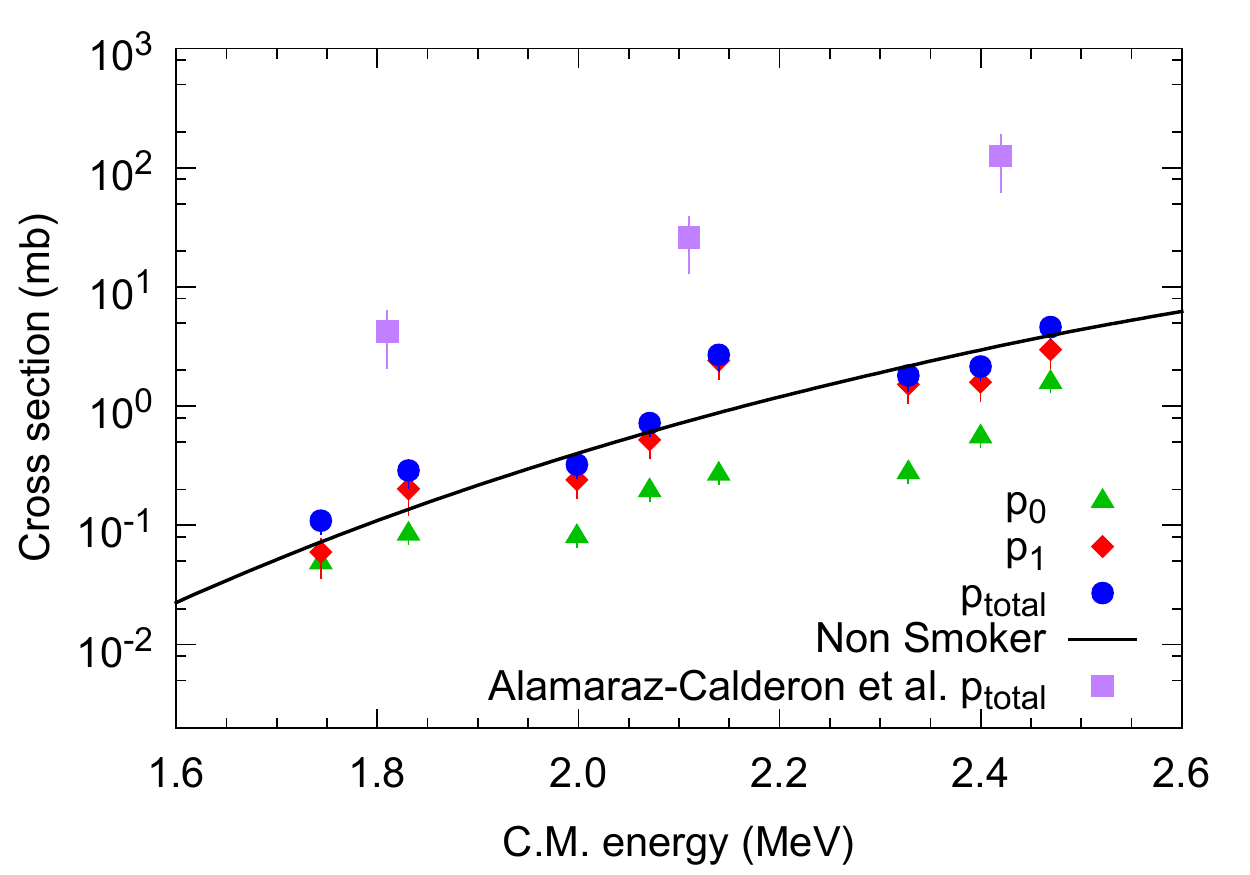}
\caption{\label{fig:compare} Cross sections for the reaction $^{23}$Na($\alpha$,$p$)$^{26}$Mg. The energies given are effective energies, corrected for energy losses within the target. See the text for details. For comparison cross sections from the statistical model code Non Smoker \cite{RAUSCHER200147} and the measurement reported in Ref.~\cite{PhysRevLett.112.152701} are also shown.}
\end{figure}

Total cross sections for $p_0$ and $p_1$ are presented in Table~\ref{tab:sigmas}. These values are plotted in Fig. \ref{fig:compare} together with results from Ref.~\cite{PhysRevLett.112.152701} and the statistical model code Non Smoker \cite{RAUSCHER200147}. We find a significant discrepancy with the results reported in Ref.~\cite{PhysRevLett.112.152701}, these values being consistently an order of magnitude higher than measured here. We can offer no explanation for this discrepancy, however it cannot be accounted for by the form of the angular distributions assumed in Ref.~\cite{PhysRevLett.112.152701}, where data were only obtained backwards of $\theta_{\text{c.m.}}=160^{\circ}$. In the narrow angular range between $\theta_{\text{c.m.}}=165^{\circ}$ and $170^{\circ}$ where overlapping differential cross section measurements exist, the absolute values again differ by at least an order of magnitude. It is again worth noting that the absolute normalisation in the present work is provided by Rutherford scattered beam from the $^{23}$Na component of the target itself. Combined with the relative simplicity of the experimental setup, this provides an extremely robust method for the determination of absolute cross sections.

\begin{table}
\caption{\label{tab:sigmas} Angle-integrated cross sections for the $p_0$ and $p_1$ branches of the $^{23}$Na($\alpha$,$p$)$^{26}$Mg reaction. The final column gives the ratio of the measured cross section to that calculated using the statistical model code Non Smoker.}
\begin{ruledtabular}
\begin{tabular}{cccc}
\textrm{E$_\text{c.m.}$ (keV)\footnote{Effective energy corrected for energy loss within the target. See the text for details.}}&
\textrm{$\sigma$ $p_0$ (mb)}&
\textrm{$\sigma$ $p_1$ (mb)}&
\textrm{$\sigma(p_0+p_1)$/$\sigma_{N.S.}$}\\
\colrule
1744	&	0.05	(1)	&	0.06	($^{+2}_{-2}$)	&	1.50	(29)	\\
1831	&	0.09	(2)	&	0.20	($^{+6}_{-7}$)	&	2.09	(46)	\\
1998	&	0.08	(2)	&	0.24	($^{+5}_{-5}$)	&	0.81	(13)	\\
2071	&	0.20	(4)	&	0.52	($^{+11}_{-12}$)	&	1.19	(19)	\\
2139	&	0.28	(6)	&	2.42	($^{+49}_{-53}$)	&	3.20	(58)	\\
2328	&	0.28	(6)	&	1.52	($^{+31}_{-34}$)	&	0.84	(14)	\\
2400	&	0.57	(11)	&	1.59	($^{+32}_{-35}$)	&	0.73	(11)	\\
2469	&	1.62	(33)	&	2.97	($^{+60}_{-66}$)	&	1.18	(17)	\\
\end{tabular}
\end{ruledtabular}
\end{table}

The Non Smoker results reproduce the measured cross sections extremely well in terms of both trend and magnitude. The only significant deviation is found at E$_{\text{c.m.}}=2.16$~MeV and can be understood in terms of the strong individual resonance reported in Ref.~\cite{PhysRevC.9.996} at E$_{\text{c.m.}}=2.14$~MeV. If the energy dependence of the Non Smoker results is fixed and only the absolute magnitude allowed to vary we find a scaling factor of 0.96$\pm$0.06 is required to best fit our data.

In conclusion, we have presented cross sections for the $^{23}$Na($\alpha$,$p$)$^{26}$Mg reaction in the region $E_{c.m.}$ = 1.74 to 2.47 MeV. The overall trend and magnitude of the cross section are in general found to be very well reproduced by the statistical model code Non Smoker. The results are also largely consistent with the previous measurements of   Whitmire \textit{et al.} \cite{PhysRevC.9.996} and Kuperus \textit{et al.}  \cite{Kuperus19642253}, though in general slightly higher than their results, whereas our measurement is inconsistent with the recent measurement by Almaraz-Calderon et al. \cite{PhysRevLett.112.152701}.

\begin{figure}[h]
\includegraphics[width=80mm]{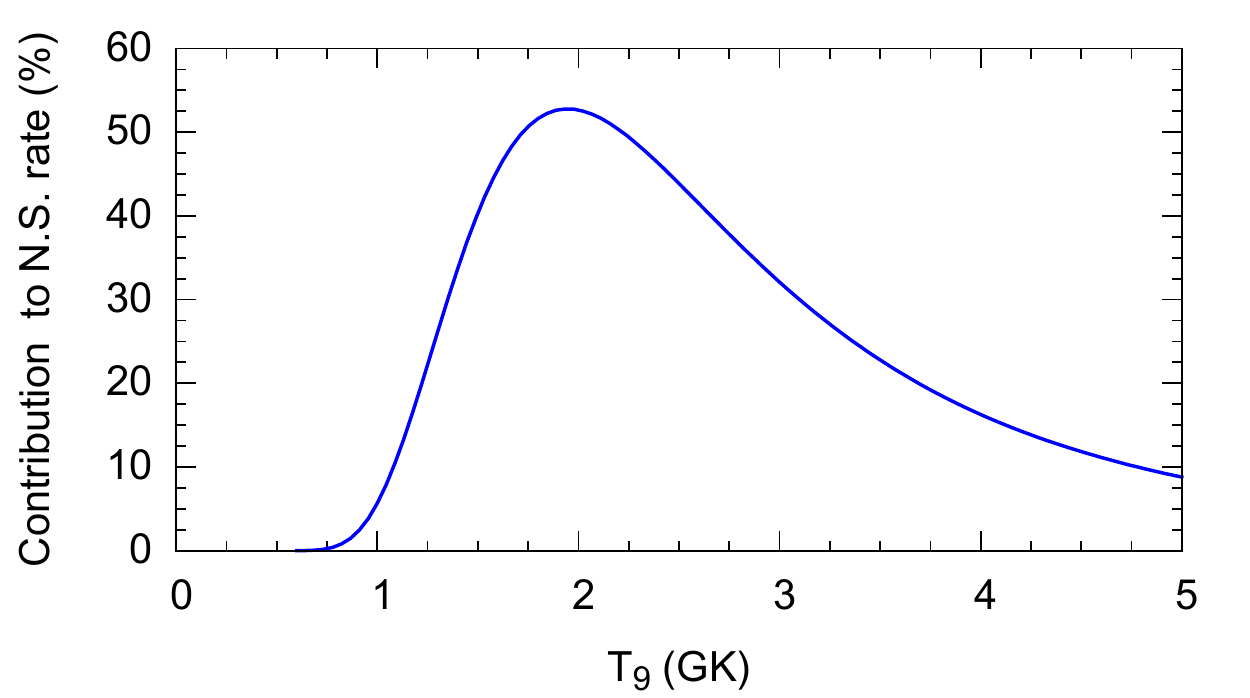}
\caption{\label{fig:resonance} The single-resonance contribution to the total rate obtained from Non-Smoker calculations based upon the measured strength of the resonance at $E_{c.m.} = 2.14$~MeV (see text for details).} 
\end{figure}

As mentioned, the only significant discrepancy between the Non-Smoker statistical model and our measurement is at the energy of the strongest ($\alpha$,$p$) resonance at $E_{c.m.} = 2.14$~MeV, a resonance which is particularly strong in the $p_1$ channel. From the difference between the observed cross sections around 2.07~MeV and 2.14~MeV centre of mass energy, we estimate the $p_1$ and $p_0$ resonance strengths for this resonance to be $\omega\gamma_1$ = 1000(300)~eV and $\omega \gamma_0$ = 42(13)~eV respectively. Based on these resonance strengths, the corresponding single-resonance contribution to the reaction rate is shown in Fig. \ref{fig:resonance} compared to the Non-Smoker reaction rate. The contribution from this resonance in itself exhausts up to 50\% of the Non-Smoker reaction rate (at 2~GK), and could therefore potentially increase the total reaction rate beyond that of the Non-Smoker rate. At the most important temperature, 1.4~GK, the temperature at termination of convective shell C/Ne burning \cite{0067-0049-193-1-16}, the single-resonance contribution to the reaction rate is 35\% of the Non-Smoker reaction rate, with a reduced contribution below that temperature. Based on this, we would still recommend usage of the Non-Smoker reaction rate for $^{23}$Na($\alpha$,$p$)$^{26}$Mg reaction in astrophysical scenarios, rather than the reaction rate indicated in Ref.\cite{PhysRevLett.112.152701}. The error on the reaction rate as evaluated from our experimental data is significantly reduced to the level of 30\% relative error on the reaction rate, except in the temperature region around 2~GK where the contribution from the resonance could increase the reaction rate by up to 50\% as shown in Fig. \ref{fig:resonance}, with a corresponding increase in the upper limit on the reaction rate.

In summary, we therefore conclude that the reaction rate in the key temperature region, around 1.4~GK, is consistent with that of the statistical model (Non-Smoker), to within approximately 30\%. Based on this, the resulting $^{26}$Al production in massive stars as presented in Ref. \cite{0067-0049-193-1-16} still stands. From the results of this sensitivity study, in which a 30\% $^{26}$Al production increase is found for a rate-increase of a factor of two, the uncertainty in the $^{26}$Al production corresponding to our reaction-rate uncertainty of 30\% is expected to be at most 10--20\%. This level of precision in the $^{23}$Na($\alpha$,$p$)$^{26}$Mg reaction rate should therefore be sufficient for detailed comparisons of observed and simulated astrophysical $^{26}$Al production.

\begin{acknowledgments}
The authors would like to thank Folmer Lyckegaard for preparation of the NaCl targets used in this work. We also acknowledge financial support from the European Research Council under ERC starting grant “LOBENA”, No. 307447 and from the UK Science \& Technology Facilities Council under grant number ST/J000124/1.
\end{acknowledgments}

\bibliography{Na_ap}

\end{document}